\documentclass[prl,preprint,amsmath,amssymb,superscriptaddress,
showpacs]{revtex4} 

\usepackage{graphicx} \usepackage{color} \usepackage{dcolumn} 
\usepackage{units} 

\begin{document}

\title{Dependence of river network scaling on initial conditions} 
\author{Geoffrey M.\ Poore} \email{gpoore@illinois.edu} 
\affiliation{Department of Physics, University of Illinois, Urbana, 
Illinois 61801} \author{Susan W.\ Kieffer} \affiliation{Department of 
Geology, University of Illinois, Urbana, Illinois 61801} 
\date{September 11, 2009}
\begin{abstract} 
We investigate the dependence of river network scaling on the relative 
dominance of slope vs.\ noise in initial conditions, using an erosion 
model. Increasing slope causes network patterns to transition from 
dendritic to parallel and results in a breakdown of simple power-law 
scaling. This provides an example of how natural deviations from scaling 
might originate. Similar network patterns in leaves suggest such 
deviations may be widespread. Simple power-law scaling in river networks 
may require a combination of dynamics, initial conditions, and 
perturbations over time.
\end{abstract}
\maketitle 

Branching network patterns are ubiquitous in nature, from river networks 
to lightning, from plant vascular patterns to animal circulatory 
systems. It is generally accepted that these networks exhibit power-law 
scaling. While it has been suggested that the scaling is topologically 
inevitable \cite{Paik2007}, topology may have limited relevance for 
networks embedded in physical space \cite{Dodds2000}, and in any case 
allows a range scaling exponents \cite{Paik2007}. 

The origin of specific exponent values remains a topic of active 
research. In river networks, three major approaches have been taken: 
growth models that create static networks through a growth mechanism 
\cite{Niemann2001}, optimal models that create equilibrium networks 
through an optimization process \cite{Rodriguez-Iturbe1997}, and dynamic 
models based on local erosion rules iterated over time 
\cite{Sinclair1996,Somfai1997,Whipple1999}. We are interested in how the 
scaling of dynamic models is affected by anisotropy in initial 
conditions. In a different type of network, anisotropic 
diffusion-limited aggregation, scaling exponents depend on the strength 
of anisotropy \cite{Popescu2004}. In river networks, landscape slope is 
known to affect network appearance 
\cite{Zernitz1932,Howard1967,Twidale2004}, but possible effects on 
scaling have been almost entirely ignored \cite{Mejia2008}. 

The study of scaling laws in river networks goes back to Horton's work 
in the 1940s \cite{Horton1945}. Although a number of scaling laws have 
been discovered, only two scaling exponents are independent for most 
networks \cite{Dodds1999,Dodds2000}: the Hack exponent $h$ and the 
sinuosity exponent $d$. The Hack exponent is defined by $\langle l 
\rangle\sim A^h$ and the sinuosity exponent by $\langle l \rangle\sim 
L_\parallel^d$, where $\langle l\rangle$ is average main stream length, 
$A$ is drainage area (land area draining to a point), and $L_\parallel$ 
is longitudinal (Euclidean) main stream length. The natural range for 
$h$ is $0.5$--$0.7$, with $0.57$ a commonly cited average; $d$ ranges 
from $1.0$--$1.2$, with $1.1$ a common average 
\cite{Maritan1996a,Rigon1996,Tarboton1988,Dodds2000}. The basin shape 
exponent $D=d/h$ is also useful in network studies \cite{Dodds1999}. 
Since $A\sim L_\parallel^D$, $D$ indicates whether the scaling of basin 
shape is self-similar ($D=2$) or self-affine ($D\ne 2$) 
\cite{Maritan1996a}. 

In this Letter, we investigate the dependence of river network scaling 
on initial conditions (ICs), using a dynamic model. We consider the 
effect of the relative dominance of slope vs.\ noise in the ICs, or 
equivalently, the degree to which ICs are anisotropic vs.\ isotropic. 
Anisotropy of ICs is quantified, and related to initial scaling. We find 
that anisotropy produces deviations from simple power-law scaling in 
steady-state networks, because the sensitivity of network structure to 
anisotropy is scale-dependent. This provides a simple example of how 
scaling deviations in natural rivers may originate. As ICs become more 
anisotropic, network patterns transition from dendritic to parallel. 
Similar patterns in leaf veins suggest that deviations from scaling 
exist in other networks. Since model dynamics do not guarantee simple 
power-law scaling, power laws in river networks may require a 
combination of dynamics, ICs, and perturbations that smooth scaling 
deviations. 

We simulated an erosion law on a square lattice. All nodes of the 
lattice receive uniform precipitation. Water from each node flows to the 
neighboring node with the steepest downslope gradient; diagonal flow is 
allowed. Since it is possible to have nodes with no lower neighbors 
(pits), lakes can form and a lake algorithm is needed to route pit flow. 
Our lake algorithm fills depressions with water and finds the lowest 
outlet, where the lake overflows. The algorithm also prevents erosion of 
nodes below the lake surface. The system evolves from a random initial 
surface to a static steady-state network. 

\begin{figure*} 
\begin{center} 
\includegraphics[width=\textwidth]{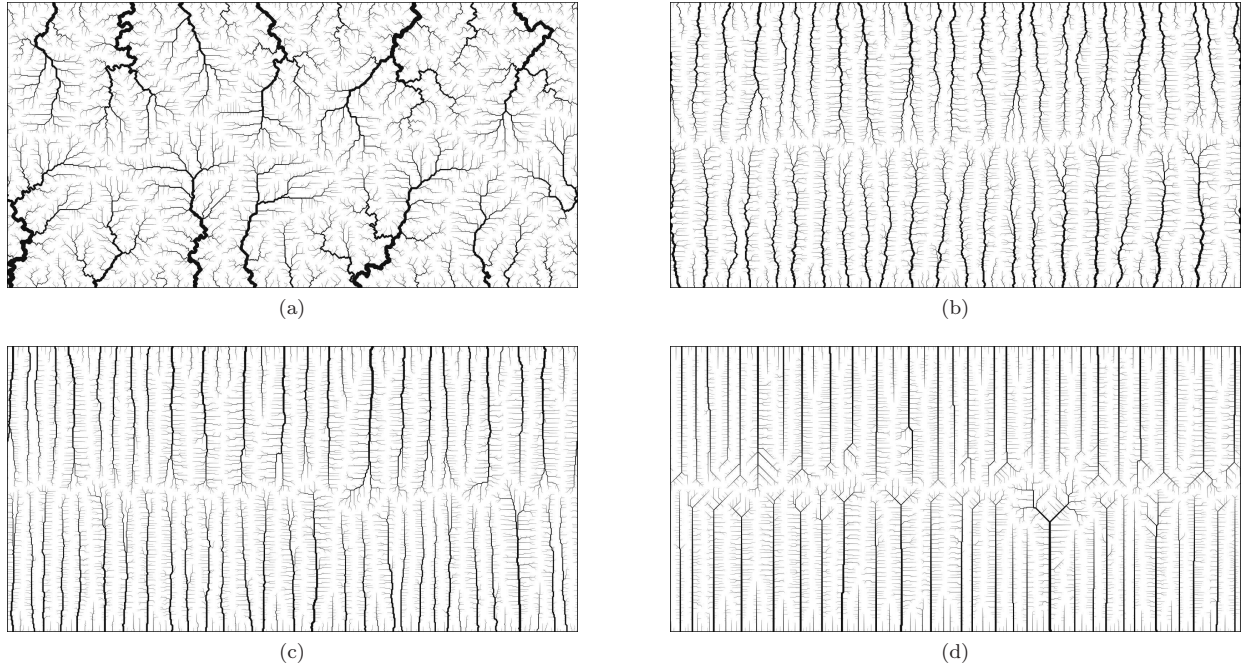} 
\end{center} 
\caption{\label{fig:netpic} Four steady-state river networks produced by 
simulation of Eq.\ \ref{eqn:erosion-differential}. Initial 
slope-to-noise ratios $\lambda$, from (a)--(d), are $0$, $1$, $2$, and 
$3$. Noise dominated the initial condition of (a). Noise and slope were 
initially balanced in (b). Initial slope dominated (c), and by (d), 
slope was so dominant that the final drainage pattern is unnaturally 
regular. The drainage pattern of (a) is dendritic, while (b)--(d) have 
parallel patterns \cite{Zernitz1932,Howard1967,Twidale2004}. Only rivers 
with drainage area $A\ge 10$ nodes are shown; line widths are 
proportional to $\sqrt{A}$.} \end{figure*} 

The simulated erosion law is of the form \begin{equation} 
\frac{dz}{dt}=-KA^m\left|\nabla z\right|^n+U 
\label{eqn:erosion-differential} \end{equation} where $z$ is elevation, 
$K$ is an erosivity coefficient (related to rock hardness, rainfall 
rate, etc.), $m$ and $n$ are constant exponents, and $U$ is tectonic 
uplift rate. This erosion law may be derived from the assumption that 
erosion is proportional to a power of shear stress or of unit stream 
power, combined with hydrodynamic principles \cite{Whipple1999}. The 
same form has also been suggested as a modified version of the 
Kardar-Parisi-Zhang (KPZ) equation \cite{Somfai1997}. We used the 
following typical literature values for the constants in Eq.\ 
\ref{eqn:erosion-differential}: $K=10^{-5}\,\text{yr}^{-1}$; 
$U=10^{-3}\,\text{m/yr}$; and $m=\nicefrac{1}{2}$ and $n=1$ (consistent 
with the empirical observation that often $m/n\approx\nicefrac{1}{2}$) 
\cite{Whipple1999}. 

Simulations were performed on lattices with a 2:1 width:height aspect 
ratio. Left-right boundary conditions were periodic; top-bottom were 
Dirichlet $z=0$. Boundary conditions were chosen to minimize edge 
effects; basin boundaries form spontaneously. 

Initial conditions consist of two sloping surfaces that meet in the 
center. Random noise in elevation from a uniform distribution is added 
to these sloping surfaces. To specify the relative dominance of slope 
vs.\ noise (anisotropy vs.\ isotropy) in the initial conditions, we 
introduce the initial slope-to-noise ratio $\lambda=|s|\Delta x/N$, 
where $s$ is the slope of the initial condition in the absence of noise, 
$\Delta x$ is the lattice spacing, and $N$ is the maximum magnitude of 
the noise in elevation (1 m for our simulations). $\lambda$ is the ratio 
of elevation change over one node from slope to the maximum possible 
change from noise. $\lambda=0$ indicates a flat surface with noise, or 
alternatively the unphysical limit where $N\rightarrow\infty$ for a 
finite slope. While the initial network for $\lambda=0$ will depend on 
how flow through lakes is defined, since no direction is favored the 
overall scaling should be similar to that of a random spanning tree, 
which has $(h,d)=(5/8,5/4)$ \cite{Dodds2000}. For $\lambda\ge 1$ the 
slope is strong enough to prevent lakes, and initial scaling will be 
that of a Scheidegger network with $(h,d)=(2/3,1)$, until $\lambda$ is 
large enough to produce non-convergent flow with the trivial exponents 
$(h,d)=(1,1)$ \cite{Dodds2000}. For $0<\lambda<1$, some lakes will be 
present, and initial scaling will exist in a crossover between 
$\lambda=0$ scaling and the Scheidegger network. 

We will now show that the combination of a square lattice and sloped 
initial conditions limits the meaningful range of $\lambda$ to 
$0\le\lambda\le \sqrt{2}/(\sqrt{2}-1)\approx 3.4$. Assume that the 
initial slope $s$, in the absence of noise, is parallel to the square 
lattice. Then the slope from a node to a diagonal neighbor will be 
$s/\sqrt{2}$. If flow is ever to be diagonal, then 
$|s|/\sqrt{2}>|s|-N/\Delta x$. That is, if noise is maximized ($N$) in 
the adjacent direction and minimized ($0$) in the diagonal direction, it 
must be sufficient to cause diagonal flow. Rearranging and substituting 
$\lambda=|s|\Delta x/N$ yields $\lambda<\sqrt{2}/(\sqrt{2}-1)\approx 
3.4$ as the condition for diagonal flow. If $\lambda>3.4$, and the 
system does not experience perturbations, flow will be non-convergent 
for the entire system evolution. Somewhat lower values of $\lambda$ are 
also problematic, because the noise is small enough that flow is 
difficult to divert, resulting in drainage patterns that are unnaturally 
regular (Fig.\ \ref{fig:netpic}). 

Fifty simulations were run for each integer and half-integer value of 
$\lambda\in[0,3]$, with lattice dimensions 800x400. All simulations were 
run to steady state, defined as no change from one timestep to the next 
within the limits of numeric precision. 

The simulated networks show a transition from dendritic to parallel 
drainage patterns with increasing $\lambda$. The $\lambda=0$ networks 
are dendritic, with a tree-like pattern composed of shorter, more 
branching streams (Fig.\ \ref{fig:netpic}a). The patterns become more 
parallel with increasing $\lambda$, with longer, straighter streams that 
more rarely intersect (Fig.\ \ref{fig:netpic}b--d). In nature, the key 
distinction between dendritic and parallel networks is the extent to 
which flow is controlled by gradient 
\cite{Zernitz1932,Howard1967,Twidale2004}, consistent with our model. 

\begin{figure} \begin{center} \includegraphics{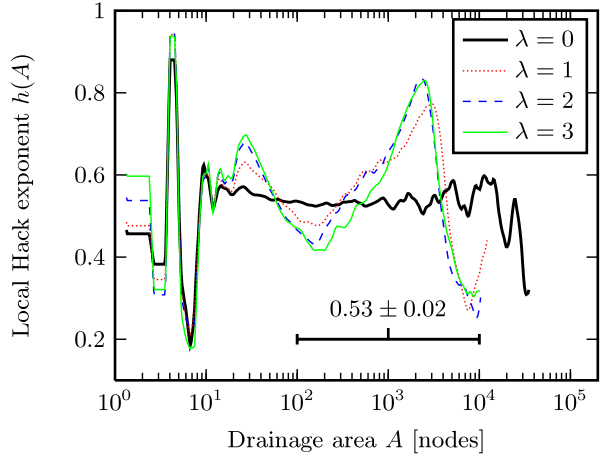} \end{center} 
\caption{\label{fig:h-vs-lambda} (Color online) The local Hack exponent 
$h(A)$ as a function of drainage area $A$ for four initial 
slope-to-noise ratios $\lambda$. Simple power-law scaling is observed 
only for $\lambda=0$. For this case, Hack's exponent $h$ was estimated 
by taking the mean and standard deviation of $h(A)$ over the region 
designated in the figure.} \end{figure} 

The simulations with $\lambda>0$ do not exhibit simple power-law 
scaling; log-log plots of $\langle l\rangle$ vs.\ $A$ and $\langle 
l\rangle$ vs.\ $L_\parallel$ do not exhibit linear behavior. To quantify 
the scaling, we introduce the local exponents \begin{equation*} 
h(A)=\frac{d\log\langle l(A)\rangle}{d\log A} \quad\text{and}\quad 
d(L_\parallel)=\frac{d\log\langle l(L_\parallel)\rangle}{d\log 
L_\parallel}. \end{equation*} The local Hack exponent $h(A)$ has been 
used previously to quantify deviations from scaling \cite{Dodds2001a}; 
local exponents have also been used in other cases where fractal 
properties are scale-dependent \cite{Chen2000,Kalda2003}. For each value 
of $\lambda$, local exponents were measured by calculating the averages 
$\langle l(A)\rangle$ and $\langle l(L_\parallel)\rangle$ over all runs, 
then performing linear regression in log-log space over a small moving 
window ($0.15$ orders of magnitude for $d(A)$ and $0.2$ for $h(A)$). For 
the figures, local exponents were smoothed by taking running averages 
over 0.05 orders of magnitude. 

The local Hack exponent $h(A)$ shows simple power-law scaling for 
$\lambda=0$ (Fig.\ \ref{fig:h-vs-lambda}). The region with 
$h\approx\text{const}$ gives $h=0.53\pm 0.02$, where the error is the 
standard deviation of $h(A)$ within the region and the values are 
calculated from unsmoothed data. There are deviations in $h(A)$ below 
$A\approx 10$ nodes due to grid effects \cite{Hergarten2001}; deviations 
due to finite size effects are observed at large $A$. For $\lambda>0$, 
no significant regions of simple scaling are observed. After increasing 
for small $A\gtrsim 10$, $h(A)$ decreases for intermediate $A$, 
increases again for large $A$, and finally decreases for the largest $A$ 
due to finite size effects. The magnitude of variation in $h(A)$ 
increases with increasing $\lambda$. 

\begin{figure} \begin{center} \includegraphics{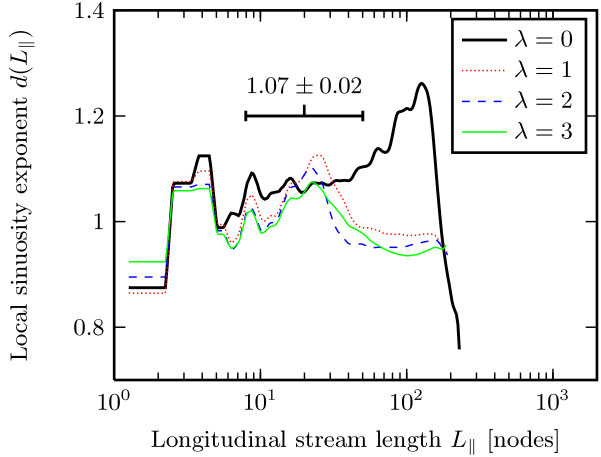} \end{center} 
\caption{\label{fig:d-vs-lambda} (Color online) The local sinuosity 
exponent $d(L_\parallel)$ as a function of longitudinal main stream 
length $L_\parallel$ for four initial slope-to-noise ratios $\lambda$. 
Approximate power-law scaling is observed for $\lambda=0$.} \end{figure}

The behavior of $d(L_\parallel)$ is similar (Fig. 
\ref{fig:d-vs-lambda}). For $\lambda=0$, there is a region of 
approximate simple scaling that gives $d=1.07\pm 0.02$. For larger 
$\lambda$, no significant regions of simple scaling are observed at 
intermediate scales. 

The breakdown of simple scaling for $\lambda>0$ may be traced to a 
scale-dependent effect of initial slope on network structure. The 
largest rivers form on the initial slope and evolve under its full 
effect. Their tributaries also initially flow down this slope, but as 
the major rivers carve out valleys, their tributaries redirect to flow 
down the valley sides. This rearrangement introduces additional noise 
into smaller-scale networks. Thus, the effect of initial anisotropy 
decays with decreasing scale, so that networks become more irregular and 
more dendritic as scale decreases. The resulting deviations from scaling 
are exacerbated by the structure of parallel networks. Between major 
junctions, length and area will increase roughly linearly 
\cite{Dodds2001a}. Since streams intersect more rarely in networks that 
are more parallel, such networks will exhibit larger fluctuations in 
their Hack distributions. 

Deviations from simple power-law scaling for Hack's law have been 
observed in continent-scale river networks \cite{Dodds2001a}. There are 
deviations at small scales due to long, thin basins, and at large scales 
due to statistical fluctuations and boundaries. At intermediate scales, 
where simple scaling would be anticipated, $h(A)$ exhibits gradual drift 
rather than fluctuating about an average value. These deviations cannot 
be attributed solely to ICs, since continent-scale networks have complex 
histories and are shaped by numerous processes, but the conditions under 
which such networks form are surely responsible for some deviations. 
More importantly, our results provide an example of how such deviations 
in river network scaling may be produced. 

The Hack and sinuosity exponents have not been measured for different 
drainage patterns as such, but recently Mej\'{i}a and Niemann 
\cite{Mejia2008} have shown that the scaling of basin shape is 
self-similar for dendritic networks and self-affine for parallel 
networks. For our simulations, the basin shape exponent $D\approx 2$ 
(self-similar) for the dendritic network ($\lambda=0$). This suggests 
that while $h$ and $d$ evolve from the initial values, self-similarity 
is inherited from the ICs. For larger $\lambda$ (more parallel), a local 
shape exponent $D(A)$ shows similar behavior to $1/h(A)$. Self-affinity 
is not inherited from the ICs since simple scaling breaks down, but a 
narrow aspect ratio is inherited. Since the scaling of parallel networks 
reported by Mej\'{i}a and Niemann shows larger fluctuations than that of 
dendritic networks, it is possible that they actually measured an 
average over local scaling rather than simple scaling. For our 
simulations, an averaged $D(A)$ would be less than $2$ and thus 
self-affine. 

It is also possible that at least some natural parallel networks do 
exhibit simple scaling. Such networks might result from ICs with 
$\lambda$ very close to zero, but they might also indicate shortcomings 
in our model. It has been shown that dynamic models give cleaner power 
laws when there are perturbations from variable boundary conditions. 
Steady-state solution methods for Eq.\ \ref{eqn:erosion-differential} 
that involve perturbing the network also yield cleaner power laws 
\cite{Hergarten2002}. Our results show that Eq.\ 
\ref{eqn:erosion-differential} does not guarantee even approximate power 
laws. Perhaps simple scaling in river networks results from a 
combination of ICs that set drainage patterns, dynamics that govern 
network evolution, and perturbations that smooth deviations from 
scaling. 

We have shown that as initial slope becomes more dominant, drainage 
patterns transition from dendritic to parallel, and simple power-law 
scaling is replaced by scale-dependent exponents. Initial conditions 
deserve greater consideration in river network models, especially given 
the range of initial conditions in use 
\cite{Sinclair1996,Braun1997,Somfai1997,Passalacqua2006,Perron2008}. 
While our results challenge attempts to explain river network scaling in 
terms of a single principle or universality class, they also open a 
number of avenues for future research, such as the effect of other 
initial geometries and other sources of anisotropy and noise. In 
particular, perturbations over time may be important in producing simple 
scaling. Our results may be relevant to other branching patterns as 
well. For example, the vein structure of some leaves 
\cite{Pelletier2000} resembles our parallel river networks, with a more 
ordered structure at large scales transitioning to a more disordered 
structure at small scales, hinting that other networks may exhibit 
complex scaling. More broadly, Chen and Bak have suggested that 
length-scale-dependent scaling ``may represent a quite general 
geometrical form for nonequilibrium dissipative systems'' 
\cite{Chen2000}. 

\acknowledgments{This research was supported by the Charles R.\ 
Walgreen, Jr., endowment to SWK. It was partially supported by the 
National Center for Supercomputing Applications under grant number 
TG-EAR080025 and utilized the SGI Altix cluster.}

\end{document}